\documentclass[12pt]{svmult}
\usepackage{graphicx}
\usepackage{makeidx}
\usepackage{multicol}

\begin{document}

\title*{A Nonequilibrium Lattice Gas of Two-species: Monte Carlo Investigations}
\author{Edward Lyman and B. Schmittmann}
\institute{Center for Stochastic Processes in Science and Engineering, and
Physics Department, Virginia Tech, Blacksburg, VA 24061-0435, USA}
\maketitle

\vspace{-.5cm}

\begin{abstract}
We present the phase diagram of a far from equilibrium system, mapped by
Monte Carlo simulation. The model is a lattice gas of two species and holes.
The two species are biased to hop in opposite directions and interact via
excluded volume and nearest neighbor attractions. Three phases are found as
function of temperature and charge density.
\end{abstract}

\vspace{-.3cm}

\emph{Introduction. }A comprehensive theory of nonequilibrium 
steady states is not yet available. This motivates the study
of simple models for which analytic results can be compared to simulation
data. While such models may seem far removed from physical phenomena, it is
clearly easier to develop the tools for treating nonequilibrium systems in
such a controlled setting before turning to the more demanding task of
describing systems found in the natural world. For example, it is not yet
known to what extent the idea of universality can be applied to
nonequilibrium systems, or for that matter how to rigorously define a phase
transition far from equilibrium. For the time being,
we deal with these questions intuitively, amassing detailed descriptions of
nonequilibrium systems while awaiting a general theory. In this paper we
present the results of a Monte Carlo (MC) study of an interacting lattice
gas driven far from equilibrium. In the first part we review the earlier
work which motivates our study, and in the second part we describe the
microscopic model and order parameters. Finally we describe the main results
of our study, a phase diagram\cite{LS} and a detailed finite-size scaling
analysis of a particular limit of the model.

Twenty years ago, Katz \textit{et al.} \cite{KLS} introduced a simple
modification (the ``KLS model") of the attractive Ising lattice gas \cite{ising}, in
which particle-hole exchanges along a particular lattice direction (which we call $y$)
are coupled to a bias $\left( E\right) $ which favors (suppresses) 
particle moves in the $+y$ ($-y$) direction. With the addition of
periodic boundary conditions in the $y$ direction, a nonzero particle
current is established and the steady state distribution is not proportional
to the Boltzmann distribution. A review of this work can be found in 
\cite{DDSrev}. 
Simulations\cite{ITAL prepr},\cite{Wang},\cite{Leung} 
and field-theoretic renormalization group studies%
\cite{FT} indicate that the model retains a continuous transition into a
low-temperature phase separated state, though the universal behavior is
distinct from the Ising class (see below). Further, in contrast to
the Ising model, the hopping
bias allows only interfaces \emph{parallel} to $y$.

Now consider a generalization of such driven Ising models \cite{SHZ} 
to two species of particles which react in opposite senses to the bias. One
type ($+$) is pushed in the $+y$ direction while the other ($-$) is pushed
in the $-y$ direction. In the limit $E,T\rightarrow \infty ,$ while $%
E/T\equiv \widetilde{E}$ finite the only interaction between particles is
excluded volume. In this case a phase transition is found as the mass
density $\left( m\right) $ and $\widetilde{E}$ are varied. At low $m$ and $%
\widetilde{E}$ the system is disordered. At sufficiently high $m$ and $%
\widetilde{E}$, the two species lock into a 'traffic jam': each blocks the
other, and one observes a high-density strip \emph{perpendicular} to $y$.

If the high $T,\widetilde{E}$ constraint is lifted in the two-species model,
then the attractive interactions will become significant over some range of $%
T$. This introduces the interesting possibility of transitions between the
two types of order mentioned above. Imagine, e.g., beginning with the KLS model
at half-filling. Now stay at half filling, but lower the charge
density, i.e., change a few `$+$' particles into `$-$'. At some critical
charge density, $q_{c}\left( E\right) $, the blocking transition may become
stable. This possibility is investigated in the following sections by
mapping the parameter space with MC simulations. Along the way we will find that
it is quite difficult to find $q_{c}\left( E\right) $, as we lack a detailed
understanding of the appropriate scaling forms in this region of parameter
space. The subtleties of scaling arguments far from equilibrium will be
illustrated in the last section, where we will determine $T_{c}$ for the KLS 
model at \emph{finite} $E$.

\emph{Microscopic Model and Order Parameters. }A configuration of the model
is specified by a set of occupation variables, $\left\{ s\left( \mathbf{r}%
\right) \right\} $, where $\mathbf{r}\equiv (x,y)$ labels a site on a fully
periodic square lattice of dimensions $L_{x}\times L_{y}$, and each $s\left( 
\mathbf{r}\right) $ can take the values $+1$, $-1$, or $0$ for a positive
particle, negative particle, or hole. We also introduce the mass variable $%
n\left( \mathbf{r}\right) \equiv \left| s\left( \mathbf{r}\right) \right| $.
We will remain always at half-filling. The charge density is defined as $q=%
\frac{1}{L_{x}L_{y}}\sum_{\mathbf{r}}s\left( \mathbf{r}\right) $. All
particles interact via the usual Ising Hamiltonian, $H=-4J\sum_{\mathbf{r},%
\mathbf{r}^{\prime }}n\left( \mathbf{r}\right) n\left( \mathbf{r}^{\prime
}\right) $, \emph{independent} of charge, with $J>0$ and the sum over
nearest neighbors. $J=1$ is chosen arbitrarily: this merely sets an energy
scale. The bias, $E$, points in the positive $y$-direction and is measured
in units of $J$. A configuration evolves by selecting a nearest-neighbor
bond at random; if occupied by a particle-hole pair, its contents are
exchanged according to the Metropolis \cite{MRRTT} rate $\min \left\{ 1,\exp
[-(\Delta H-\delta yEs(\mathbf{r)})/T]\right\} $. The second term models the
effect of the drive: if the particle, of charge $s$, is initially located at 
$\mathbf{r}$, $\delta y$ is the change in its $y$-coordinate due to the
jump. Thus, positive (negative) charges jump preferentially along (against)
the field direction. The parameter $T$ (``temperature'') models the coupling
to a thermal bath. The natural control parameters for our study are
temperature $T$ (measured in units of the Onsager value), the drive $E$ and
the charge density $q$.

Conservation laws (for $q$ and $m$) ensure spatially inhomogeneous ordered
phases. We therefore select an order parameter sensitive to such structures,
i.e., the equal-time structure factor associated with the particle
distribution $\left\langle \Phi (m_{x},m_{y})\right\rangle \equiv
\left\langle \left| \frac{\pi }{L_{x}L_{y}}\sum_{\mathbf{r}}n(\mathbf{r})e^{i%
\mathbf{k\cdot r}}\right| ^{2}\right\rangle $, $\mathbf{k}\equiv 2\pi (\frac{%
m_{x}}{L_{x}},\frac{m_{y}}{L_{y}})$. $\left\langle \cdot \right\rangle $
denotes a configurational average, and the integers $m_{x},m_{y}$ index the
wave vector. For a perfect strip along the $y$-direction, $\left\langle \Phi
(1,0)\right\rangle =1$ while a random configuration gives $\left\langle \Phi
\right\rangle =O(\frac{1}{L_{x}L_{y}})$. Except where noted, all simulations
are run on $40\times 40$ lattices, starting from random initial
configurations. One MC step (MCS) is defined as $2L_{x}L_{y}$ update
attempts. The first $2\times 10^{5}$ MCS are discarded, and measurements are
taken every $200$ MCS for the next $8\times 10^{5}$ MCS.

\emph{Monte Carlo Results: Phase Diagram. }Although most detailed studies of
the KLS model have considered the limit of infinite $E$ to accentuate
nonequilibrium properties, we are forced to consider finite values of $E$.
As our two-species simulation stumbles through phase space it may find
itself in a blocked configuration in a region of parameter space where such
states should only be metastable. At large values of $E$ this metastable
configuration will persist far beyond the time of our simulation. Initial
runs indicated that metastable lifetimes at $E=2$ are reasonable. Note that
at this value, jumps favoring $E$ provide only \emph{one-half} the change in
energy required to break a \emph{single} nearest-neighbor bond. It is
therefore natural to wonder whether the transition at such small $E$ remains
in the KLS universality class. This question will be answered in the
affirmative in the next two sections.

\begin{figure}[tbp]
\input{epsf}
\begin{center}
\begin{minipage}{0.8\textwidth}
  \epsfxsize = \textwidth \epsfysize = .8\textwidth \hfill
  \epsfbox{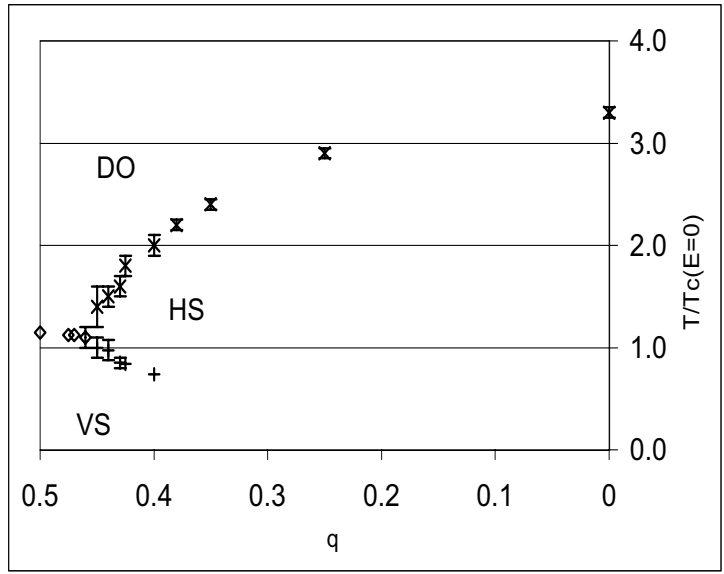}
\end{minipage}
\end{center}
\caption{Phase diagram for $E=2$ in charge density $q$ and temperature $T$, in units
of $T_{c}\left( E=0\right) $. $\left( +\right) $ denotes the first-order
line, while the diamonds and x's  denote the
continuous lines. Configurations labelled: DO for disordered, HS for horizontal strip,
VS for vertical strip} 
\vspace{-0.5cm}
\end{figure}

Fig. 1 shows the phase diagram in $q$ and $T$ for $E=2$. The phase boundaries
are mapped by sweeping in $T$ at fixed $q$ while monitoring the order
parameters $\left\langle \Phi (1,0)\right\rangle $ and $\left\langle \Phi
(0,1)\right\rangle $ and their fluctuations. The far left point is the KLS
transition at finite $E$. Moving right, parallel to the horizontal axis, we
are changing $+$ particles into $-$; as long as the $-$ are insufficient to
form a blockade the transition from disorder into a parallel (``vertical") strip
apparently remains continuous, as indicated by a peak in the
fluctuations of $\left\langle \Phi (1,0)\right\rangle $. We suspect that
this transition can be described by the KLS field theory, with the
additional complication of a small concentration of randomly distributed
(but annealed) impurities. At the far right of the phase diagram the system
contains equal numbers of $+$ and $-$ particles. There, we encounter a
transition into a blocked phase (``horizontal strip") which also appears continuous, 
indicated
here by a peak in the fluctuations of $\left\langle \Phi (0,1)\right\rangle $.
At larger values of $q$ and lower $T$, we are able to
observe a transition from the blocked phase into a vertical strip phase.
This transition appears to be first-order, displaying hysteresis and
metastability. The most interesting region in the phase diagram is where the
three lines join at $q_{c}(E)$. 
Future studies will focus on scaling properties in the vicinity of this 
nonequilibrium bicritical point. 

\emph{Anisotropic Finite-Size Scaling. }We now turn to a detailed discussion
of the KLS transition at finite $E$. As mentioned in the introduction, a
Langevin equation for a mesoscopic version of the local spin variable has
been studied in great detail\cite{FT}. The most remarkable prediction of the
field theory is a nontrivial anisotropy exponent, $\Delta>1$, so that 
wavevectors scale as $k_{\Vert }\sim k_{\bot
}^{1+\Delta }$; physically this implies that domains of correlated spins
grow faster in the field direction. In order to control finite size 
corrections, it is then necessary to account for this anisotropy \cite{ITAL prepr}. 
The anisotropy
introduces \emph{different} correlation length exponents parallel($\nu
_{\Vert }$) and perpendicular($\nu _{\bot }$) to the drive, an effect which
we will refer to as \emph{strong anisotropy}, in contrast with, e.g., an
Ising model with anisotropic interactions. The values of the critical
exponents are known from an RG analysis to all orders; their values in $d=2$
are $\beta =1/2,\nu _{\Vert }=3/2,\nu _{\bot }=1/2,\Delta =2$\cite{FT}.
Phenomenological scaling forms \cite{Leung} involve two length scales, 
$L_{\Vert }$ and $L_{\bot }$, so that the order parameter scales as
$m\left( t,L_{\Vert },L_{\bot }\right) =L_{\Vert }^{-\beta /\nu
_{\Vert }}\overline{m}\left( tL_{\Vert }^{1/\nu _{\Vert }},L_{\Vert }^{\nu
_{\bot }/\nu _{\Vert }}L_{\bot }^{-1}\right) $ and the scaling function
depends on a ''shape factor,'' $S\equiv L_{\Vert }^{\nu _{\bot }/\nu _{\Vert
}}L_{\bot }^{-1}$\cite{Leung},\cite{Wang}. Increasing the system
size while holding $S$ fixed allows us to approach $T_{c}$ without cutting 
off parallel correlations before transverse ones. We then 
use the (predicted) exponents to analyze our data. The
validity of this approach will be judged by the quality of data collapse for 
$m$. In this way we will determine $T_{c}\left( E=2\right) $, as it is the
only fit parameter. Detailed work on carefully defined correlation functions
and lengths \cite{ITAL prepr} is forthcoming.

\emph{Monte Carlo Results for E=2. } We choose as our order parameter $%
m\equiv \left\langle \left| \frac{\pi }{L_{x}L_{y}}\sum_{x,y}n(x,y)e^{2\pi
i(m_{x}x/L_{x})}\right| \right\rangle $ since it is subject to smaller
fluctuations than the structure factor. Most runs last for $1.2\times 10^{6}$
MCS, though in larger systems near criticality runs of $4.8\times 10^{6}$
MCS were needed to ensure good statistics. The first $0.2\times 10^{6}$ MCS
were discarded and measurements were taken every 400 MCS thereafter.

\begin{figure}[tbp]
\input{epsf}
\begin{center}
\begin{minipage}{0.8\textwidth}
  \epsfxsize = \textwidth \epsfysize = .8\textwidth \hfill
  \epsfbox{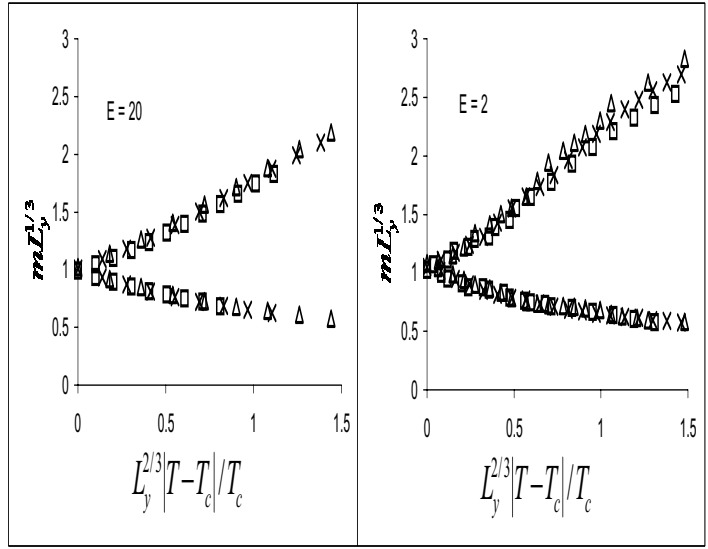}
\end{minipage}
\end{center}
\caption{Scaling of $m$ at fixed $S=.1575$ for three system sizes: $24\times 54$ (squares),
$28\times 86 \left( \times \right) ,32\times 128$ (triangles). The left (right) plot shows
$E=20$ ($E=2$). The upper (lower) branch corresponds to $T<T_{c}$ ($T>T_{c}$).} 
\vspace{-0.4cm}
\end{figure}

Fig. 2 shows the scaling of the order parameter for three different system
sizes with $S=.1575$, for both $E=20$ (effectively infinite) and $E=2$. From
these plots we estimate $T_{c}\left( E=2\right) =1.20(2)$. Notice the
systematic deviations from scaling in the $T<T_{c}$ branch, which may be due
to a small critical region or corrections to scaling from the marginal
operator. Though we have not investigated this anomaly in detail, it occurs
in the $E=\infty $ model as well\cite{Wang} and has been observed in other
nonequilibrium Ising models\cite{Racz}. Elsewhere, the data collapse is of
the same quality as in the $E=\infty $ case. We therefore have no reason to
believe that the finite $E$ transition falls into a different universality
class.

\emph{Conclusions. }We have mapped out a slice of phase space for an
interacting lattice gas of two species, driven far from equilibrium by a
bias which drives a particle current. The phase diagram has two continuous
lines which meet a first-order line at a critical charge density $%
q_{c}\left( E\right) $. In order to make more definitive claims, we need
knowledge of scaling forms in the vicinity of the bicritical point. 
It would be quite interesting to investigate this scaling behavior in order to
learn about such points far from equilibrium.

\emph {Acknowledgements.} We thank R.K.P.~Zia, and U.C.~T\"{a}uber for helpful discussions. Partial support from the National Science Foundation through DMR-0088451 is gratefully acknowledged.

\end{document}